\documentclass
[preprint,prl,showpacs,superscriptaddress,byrevtex,onecolumn]{revtex4}%
\usepackage{amsfonts}
\usepackage{amsmath}
\usepackage{amssymb}
\usepackage{graphicx}%
\begin{document}
\author{Despina Louca}
\affiliation{University of Virginia, Department of Physics, Charlottesville, VA 22904. $\ $}
\author{J. L. Sarrao}
\affiliation{Los Alamos National Laboratory, Los Alamos, NM 87545.}
\date{\today }
\title{Dynamical disorder of spin-induced Jahn-Teller orbitals with the
insulator-metal transition in cobaltates}

\begin{abstract}
Using elastic and inelastic neutron scattering, we investigated the evolution
of the local atomic structure and lattice dynamics of La$_{1-x}$Sr$_{x}%
$CoO$_{3}$ ($x=0-0.5$) as it crosses over with \textit{x} from an insulator to
a ferromagnetic metal (FMM). \ Our pair density function analysis indicates
that in the paramagnetic insulating (PMI) phase for all \textit{x}, spin
activation of Co$^{3+}$ ions induces local static Jahn-Teller (JT)
distortions.\ The size of the JT lattice increases almost linearly with
$\mathit{x.}$ \ However, in the FMM phase, static JT distortions are absent
for x $\leqslant~$30 \%. \ This coincides with narrowing of $\hbar\omega$=22
and 24 meV modes in the phonon spectrum in which we argue is due to localized
dynamical JT fluctuations. For x $>$ 30\%, static JT distortions reappear
along with broadening of the phonon modes because of weakened charge-lattice
interactions. \ \ \ \ 

\end{abstract}
\maketitle

Jahn-Teller (JT) interactions in transition metal perovskites have often led
to quite interesting physical phenomena through the interplay of orbital and
spin degrees of freedom with the lattice [1]. \ Understanding the
characteristics of JT instability can help elucidate the nature of
insulator-metal transitions (IMT), magnetoresistance and even
superconductivity in a variety of systems [2]. \ It appears that in pure
systems such as LaCoO$_{3}$ [3], LaMnO$_{3}$[4] and YTiO$_{3}$ [5],
cooperative JT or orbital ordering effects can exist up to very high
temperatures. \ LaCoO$_{3}$ is distinctly different from other perovskite
crystals because orbital degeneracy is created by thermally inducing a spin
transition [6] subjecting the system to variable spin-lattice coupling
depending on the configuration [7]. \ The ground state of Co$^{3+}$ ions at
the octahedral sites is the low spin configuration (LS), S=0 $\left(
t_{2g}^{6}e_{g}^{0}\right)  $, and this can be thermally excited to either a
high-spin (HS) $\left(  t_{2g}^{4}e_{g}^{2}\right)  $ (S=2) or an
intermediate-spin (IS) $\left(  t_{2g}^{5}e_{g}^{1}\right)  $ (S=1)
configuration [8]. \ Previous experimental [7,9] and theoretical [10] works
showed that below room temperature the IS state is more stable than the HS
because activation to this state invokes stronger oxygen $p$ to cobalt $d$
hybridization. The IS is distinct from the other two because it is $e_{g}-$JT
active: partial e$_{g}$ occupancy induces orbital degeneracy which is lifted
by changing the local crystal symmetry resulting in JT modes. Although in an
ideal cubic perovskite the crystal field energy is large [11], strong Co-O
covalency reduces the splitting between the $t_{2g}$ and $e_{g}$ bands that
enables the transition to IS. In addition, the LS-IS transition might
instigate dynamic orbital ordering [3, 10]. However, it is still unclear
whether or not ordering, either static or dynamic, occurs when charge carriers
are introduced to drive the system through an IMT. \ \ 

The addition of charge carriers establishes the FMM state by x $\geqslant
18\%~$in La$_{1-x}$Sr$_{x}$CoO$_{3}$ [12], but little is known on the
evolution of the spin transitions or the dependence of JT ordering to carrier
mobility. \ Experimental evidence suggests that the IS-JT\ state is stabilized
with doping at least in the PMI phase [7,13,14]. \ In the FMM, it would be
conceivable that charge dynamics interfere with orbital JT ordering giving
rise to a new state. \ When charge mobility increases, the JT distortions can
lose their long-range coherency. \ But they might also fluctuate dynamically
possibly giving rise to dynamical orbital ordering. \ Such dynamical JT modes
can manifest themselves as real-space modulations of orbital waves or orbitons
as suggested by Saitoh et al. [15] for LaMnO$_{3}$. \ At the same time, such
modes can also couple to the lattice forming polarons [16] or JT-polarons [17]
depending on the coupling strength. \ On the other hand, in the absence of any
ordering, a disordered structure state would emerge, creating an orbital
liquid state in an analogous way to geometrically frustrated magnetic systems
[18]. \ In this paper we report on how the local atomic structure and lattice
dynamics change from the PMI to the FMM phases in doped LaCoO$_{3}~$in support
of JT dynamical orbital disorder.

We investigated La$_{1-x}$Sr$_{x}$CoO$_{3}$ using time-of-flight pulsed
neutron scattering. \ The time-averaged structure function, S(Q($\omega=0)$,
and the dynamical, time-resolved S(Q,$\omega$) are combined to provide
information on the type of atomic distortions and how they fluctuate in time.
\ Characteristic energies of the fluctuations are defined that help explain
the crossover from static to dynamic lattice effects. \ Our measurements
indicate that lattice dynamics change continuously through the IMT both as a
function of temperature and carrier concentration. \ Below the percolation
limit, x $\leqslant$ .3, the phase transition results from a break from an
orbitally ordered, inferred from the presence of static JT in the PMI, to a JT
disordered state due to dynamical fluctuations coupled to the increase in
charge mobility. \ While the short-range local atomic structure sensitively
follows the disappearance of ordering of JT distortions close to the phase
boundary, the dynamical structure function shows that low-energy cobalt phonon
modes get narrower in energy with cooling. \ The fact that JT disorder can
create a local phonon resonance, $\hbar\omega~$= 22 meV, might be a
manifestation of a phonon mediated charge transfer mechanism that plays an
important role in the transition process of La$_{1-x}$Sr$_{x}$CoO$_{3}$.
\ Above the percolation concentration, connectivity between FM clusters is
established that allows easy charge propagation accompanied by the
reappearance of static JT and smearing of the phonon modes. \ In this case the
lattice and charge dynamics are decoupled.

The powder samples were prepared by standard solid-state reaction method
starting from SrCO$_{3}$, La$_{2}$O$_{3}$ and Co$_{3}$O$_{4}$, fired at 1100
$^{o}$C with a final firing at 1300 $^{o}$C in air, and additional annealing
in nitrogen at 900 $^{o}$C to ensure stoichiometric oxygen content for the
pure sample. \ The neutron data were collected at the Intense Pulsed Neutron
Source (IPNS). \ Data for the local structure analysis were obtained at the
Glass Liquid and Amorphous Diffractometer (GLAD). \ The total S(Q), integrated
over a finite energy window and corrected for instrumental background,
absorption, incoherent, multiple, and inelastic scattering, was Fourier
transformed to determine the pair density function (PDF). \ The PDF provides a
real-space representation of atomic pair correlations. \ Details of this
technique can be found in ref. [19, 20]. The inelastic measurements were
performed at the Low Resolution Medium Energy Chopper Spectrometer (LRMECS).
\ The incoming energy was set at E$_{i}$ = 35 meV. \ Corrections for
background and detector efficiency, empty aluminum can with and without
cadmium shielding and vanadium standard measurements were performed. The data
were also corrected for the sample self-shielding as well as the $\frac{k_{f}%
}{k_{i}}$ phase space factor. \ The data represent the sum over all scattering
angles multiplied by the Bose population factor. \ 

The temperature dependence of the PDF corresponding to the local atomic
structure for $x=0.1$ is shown in Fig. 1(a) in the vicinity of the Co-O
octahedra. The Co-O pair correlation appears $\sim$ 1.93 \AA  as the first
peak in the PDF since it is the shortest distance in the crystal. In the PMI
phase, this peak is split and a second peak appears as a shoulder to the right
$\sim$2.10 \AA  corresponding to long Co-O bonds. \ The presence of the small
second peak is attributed to static JT distortions, similarly observed in
LaMnO$_{3}$ [20], a well known JT system. In response to this local symmetry
breaking, oxygen $p$ and cobalt $d$ hybridization is stabilized by minimizing
Coulomb repulsions. \ Consequently, the crystal structure is expected to
deviate from cubic symmetry because the\ Co-O octahedra elongate in the
direction of the occupied orbital [7]. \ Below the spin-glass transition
however (T$_{SG}$ = 45 K), the peak corresponding to the long bonds vanishes
while the main Co-O peak becomes taller. \ In Fig. 1(b) the presence of the
long (LB) and short (SB) bonds is marked over an expanded temperature range.
\ They are simultaneously present at temperatures above the transition while
below only one type of bond is observed. \ For higher concentrations, the
local atomic structure also shows evidence for JT distortions in the PMI phase
(Fig. 2a). \ In the FMM however, JT distortions are not observed below x$_{c}$
$\sim$ 0.3 (Fig. 2b).

While it is expected that the number of IS-JT sites will be thermally
populated as in LaCoO$_{3}$, how that number varies with $x$ is not well
understood. \ An estimate of the JT population can be determined from the
integrated intensity of the first Co-O PDF peak using the expression of ref.
[21]. \ In the simplest scenario, one could assume that the addition of charge
creates LS-Co$^{4+}$ ions with all spins in the t$_{2g}$. \ Its local
structure is therefore equivalent to that for LS-Co$^{3+}$. \ Thus in the
\textit{insulating state}, since charges remain localized at Co$^{4+}$ sites,
the area under the Co-O peak should remain constant. \ The combined intensity
for LS-Co$^{3+}$ and LS-Co$^{4+}~$ions should account for the total area under
the peak. \ In an ideal octahedral environment, the Co-O bonds are all short
without JT and the coordination should be 6 corresponding to 100 \% LS. \ But
as seen from Fig. 2c, the total LS population corresponding to the number of
Co-O undistorted octahedra decreases from 100 \% with doping as strong
indication that activation to a different spin state must be occurring. Unless
we assume the presence of a third configuration, that of IS-JT, the total
intensity can never be accounted for. \ The estimated dependence of
LS-Co$^{4+}$, total LS (Co$^{4+}$+Co$^{3+}$) and IS-Co$^{3+}$ to the nominal
charge concentration are shown in Fig. 2(c) at room temperature (RT). \ The
split of the Co-O bonds is confirmed by Fig. 2(a) and 2(d) at RT, which is
additional evidence for the IS-JT. \ Why is the IS-Co$^{3+}$ activation
linear? One possibility is that localized charges on LS-Co$^{4+}$ in the
insulating phase stabilize the JT at neighboring sites creating IS-Co$^{3+}$.
\ The contributing factors to this are size effects and covalency [22]:
LS-Co$^{4+}$ is smaller and intermediate oxygens shift towards it while
stabilizing e$_{g}$ orbital occupancy on neighboring IS-Co$^{3+}$. \ The
covalency between Co$^{3+}$ and Co$^{4+}$ is optimized by the e$_{g}$ orbital
occupancy and allows hopping. Such a pair would be coupled ferromagnetically
with hopping occurring via e$_{g}$ forming a Zener polaron. \ Thus for every
Co$^{4+}$ there would be an IS-Co$^{3+}$ giving rise to the linear dependence
on $x$ (Fig. 2c). \ On the other hand, if activation to IS-Co$^{4+}$ were to
be considered, this would imply LS-Co$^{3+}$ but that would be energetically
unfavorable for e$_{g}$ hopping. \ \ 

Below the phase boundary, static JT\ distortions disappear, and at 10 K, the
local structure appears uniform with no distortions, and the long and short
Co-O bonds are indistinguishable for $x$ $\leq.3~$(Figs. 2b and 2d). \ But the
transition to the metallic state is not straight forward because of lattice
dynamics implications. \ Plots of the\ dynamic structure function, S$\left(
\omega\right)  $, summed over all momentum transfers, Q, for 0, .1, .3 and .4
are shown in Fig. 3. \ With Sr doping, the shape and relative intensity of
distinct low energy modes observed in the range of 10 to 30 meV change. \ In
the pure sample, we observed that the band at 22 meV is suppressed relative to
the 24 meV peak. \ Increasing $\mathit{x}$ to 0.1 (spin-glass), the two bands
are equal in height but not well resolved. \ But by x = 0.3, their intensity
is enhanced and they become narrow and better resolved signifying an unusual
electron-lattice interaction. \ The functions were fit by two gaussians in the
18-28 meV range (solid lines). \ The intensity exhibits a dependence on the
momentum transfer proportional to Q$^{2}$ suggesting that they are primarily
phononic in nature. Also the magnitude is indicative of many phonons being
affected at once. \ The S($\omega$) plots were compared to the isostructural
LaAlO$_{3}$ (not shown), another perovskite with the same rhombohedral
symmetry and similar tilting along the [111] plane [23] which showed evidence
for the low energy modes ($\sim$10 and 15 meV) but shifted by $\sim$ 2 meV to
higher energies because of mass differences of Co and Al. \ However, the
higher modes\ (22 and 24 meV) were absent in the aluminate which would suggest
that the octahedra are inactive in this system. \ The results on the
cobaltates differ from the IR measurements of ref. [24] that concluded that
the phonons do not change with Sr doping based upon information on zone-center
modes alone. \ For $x=.4$, static JT distortions reappear as seen in Fig. 2b
concomitant while the 22 and 24 meV modes smear out and can be fit by a single
gaussian centered at $\sim$23 meV. \ 

The development of the phonon modes particularly at 22 meV might be due to the
formation of new Co-O lattice vibrations with the IMT. \ The observed changes
are not due to a change in symmetry as the structure remains rhombohedral
(R$\overline{3}$C symmetry [13, 23]) on average at all \textit{x}. \ Even
though Sr is lighter than La, the peaks are not shifting to higher energies
suggesting that these modes are not dominated by the La/Sr density of states.
\ This is also supported by the fact that identical measurements on
LaAlO$_{3}$ showed no such band formation. \ The evolution of S$\left(
\omega\right)  $ with temperature through the IMT is shown in Fig. 4 for the
x=0.30 with T$_{C}$ $\sim$240 K. \ These bands may signify strong
charge-lattice coupling in the metallic state that is either suppressed or
smeared out in the insulating phase.\ \ The width of the peaks gets narrower
below the IMT suggesting phonon localization. \ 

If a LS-Co$^{4+}$ ion and an IS-Co$^{3+}$ ion are next to each other, by
moving an $e_{g}$ electron from Co$^{3+}$ ion to a Co$^{4+}$ ion they swap
configuration. \ Therefore they can form a resonant state by dynamically
sharing an $e_{g}$ electron. \ The shared $e_{g}$ electron can couple the
$t_{2g}$ electrons of the two ions ferromagnetically through double exchange.
\ As the hole doping is increased, such pairs will start to connect with each
other, forming a larger conducting network, and the entire lattice will change
into a ferromagnet. \ In this system, the crystal field splitting between the
$e_{g}$ state and the $t_{2g}$ state is of the order of $\Delta$ = 2 $eV$
[11]. \ In the undoped state this is large enough to overcome the Hund's
coupling and the difference in the Coulomb repulsion for placing 2 electrons
on the $t_{2g}$ and that for placing 1 electron each on the $t_{2g}$ state and
the $e_{g}$ state, and thus to drive the system to the LS\ state. \ In the
doped state, however, the sharing of the $e_{g}$ electron described above
creates a strong Co-O-Co bond and reduces the magnitude of $\Delta$, producing
soft Jahn-Teller modes that are low enough in energy to be strongly populated
by thermal phonons. \ Such a soft phonon mode could be the vehicle that
induces the IMT. \ At higher doping, beyond the site percolation limit of 0.31
for a cubic lattice [25], the local structure shows the reappearance of static
JT distortions even in the metallic state. \ This might be because the
concentration of charges is so high that propagation occurs through well
connected LS-Co$^{4+}$ sites without destroying the JT at IS-Co$^{3+}$ sites.

It is possible that the observed effects also describe frustration of a JT
lattice. \ In dilute JT crystals with a magnetic transition, a spin-glass
phase can sometimes exist [26]. \ This is brought about when the magnetic
moment is randomly distributed in space but frozen in specific directions, in
clusters, while correlation between clusters exists. Fluctuations of spins can
induce frustration of the JT lattice sites that can destroy their
cooperativeness and give rise to a spin-glass like state [27,28]. \ This also
affects the effective long-range elastic interactions between JT sites that
can fluctuate in sign and magnitude. \ Such JT-spin interactions are believed
to be quite uncommon [27] although glassy behavior is widely observed in
ferroelectrics and might be the case for La$_{1-x}$Sr$_{x}$CoO$_{3}$. \ For
comparison we note that the energy scale of the dynamics invoked by such
interactions are intermediate to those of the manganites and cuprates. \ In
manganites it is understood that the JT distortions are frozen. \ On the other
hand in the cuprates when distortions are static one obtains stripe structures
but when they are dynamic, as in the superconducting state, they are averaged
out. \ The cobaltate system's intermediate behavior makes it a prototypical
example for probing the nature of the crossover regime. \ In conclusion, the
local atomic structure combined with the dynamic structure function provided
evidence for a phonon mediated transition mechanism for the IMT in La$_{1-x}%
$Sr$_{x}$CoO$_{3}$.

The authors would like to acknowledge valuable discussions with J. B.
Goodenough, D. I. Khomskii and T. Egami. \ They also thank E. Goremychkin and
R. Osborn for their help with the LRMECS experiment and data analysis. \ Work
at the University of Virginia is supported under U. S. Department of Energy,
under contract DE-FG02-01ER45927, at the Los Alamos National Laboratory under
contract W-7405-Eng-36 and at the IPNS under contract W-31-109-Eng-38.\ 

[1] J. B. Goodenough, Mater. Res. Bull. \textbf{6}, 967 (1971).

[2] I. B. Bersuker, Chem. Rev. \textbf{101}, 1067 (2001).

[3] Y. Tokura and N. Nagaosa, Science \textbf{288}, 462 (2000).

[4] T. Mizokawa and A. Fujimori, Phys. Rev. B \textbf{51}, 12880 (1995).

[5] C. Ulrich \textit{et al.}, Phys. Rev. Lett \textbf{89}, 167202 (2002).

[6] J. B. Goodenough, J. Phys. Chem. Solids \textbf{6}, 287 (1958).

[7] D. Louca \textit{et al.}, Phys. Rev. B \textbf{60}, R10378 (1999).

[8] P. M. Raccah and J. B. Goodenough, Phys. Rev. \textbf{155}, 932 (1967).

[9] S. St\o len \textit{et al.}, Phys. Rev. B \textbf{55}, 14103 (1997); T.
Saitoh \textit{et al.}, Phys. Rev. B \textbf{56}, 1290 (1997); V. G.
Bhide\textit{\ et al.}, Phys. Rev. B \textbf{12}, 2832 (1975); M. Abbate
\textit{et al}., Phys. Rev. B \textbf{47}, 16124 (1993); K. Asai \textit{et
al.}, Phys. Rev. B \textbf{50}, 3025 (1994); M. Itoh and I. Natori, J. Phys.
Soc. Japan \textbf{64}, 970 (1995); S. Yamaguchi, Y. Okimoto, and Y. Tokura,
Phys. Rev. B \textbf{55}, R8666 (1997).

[10] M. A. Korotin \textit{et al.}, Phys. Rev. B \textbf{54}, 5309 (1996).

[11] A. R. West, \textit{Solid State Chemistry and its Applications }(John
Wiley \& Sons Ltd., Chichester, 1984), p. 306.

[12] G. H. Jonker and J. H. Van Santen, Physica (Amsterdam) \textbf{19}, 120 (1953).

[13] R. Caciuffo \textit{et al.,} Phys. Rev. B \textbf{59}, 1068 (1999).

[14] J. Wang \textit{et al.}, Phys. Rev. B \textbf{66}, 064406 (2002); S.
Tsubouchi \textit{et al.}, ibid \textbf{66},\textbf{ }052418 (2002).

[15] E. Saitoh \textit{et al.}, Nature \textbf{410}, 180 (2001).

[16] L. F. Feiner, A. M. Ole\'{s}, J. Zaanen, Phys. Rev. Lett.\textbf{ 78},
2799 (1997).

[17] D. Louca \textit{et al.}, Phys. Rev. B \textbf{56}, R8475 (1997).

[18] S. Ishihara, M. Yamanaka, and N. Nagaosa, Phys. Rev. B \textbf{56}, 686
(1997); M. V. Mostovoy and D. I. Khomskii, cond-mat/0201420 v1.

[19] B. H. Toby and T. Egami, Acta Crystallogr. A \textbf{48}, 336 (1992).

[20] D. Louca and T. Egami, Phys. Rev. B \textbf{59}, 6193 (1999).

[21] The number of Co-O pairs is obtained from N$_{Co-O}=\frac{4\pi
\left\langle b\right\rangle ^{2}%
{\displaystyle\int\nolimits_{r_{1}}^{r_{2}}}
\rho\left(  r\right)  r^{2}dr}{2c_{Co}b_{Co}b_{O}}$ where r$_{1}~$and r$_{2}%
~$include only the first peak, $\rho\left(  r\right)  $ is the PDF, $b_{Co,O}$
is the neutron scattering lengths for Co and O,
$<$%
b%
$>$
is the average scattering length and c$_{Co}$ is the concentration of Co atoms
per unit cell.

[22] T. Mizokawa \textit{et al}., Phys. Rev. B 63, 024403 (2000).

[23] H. D. Megaw and C.N.W. Darlington, Acta Crystallogr. A \textbf{31}, 161 (1975).

[24] S. Tajima \textit{et al.}, J. Phys. C: Solid State Phys. \textbf{20},
3469 (1987).

[25]R. Zallen, \textit{The Physics of Amorphous Solids} (John Wiley \&\ Sons,
\ Inc., New York, 1983) p. 170.

[26] F. Mehran and K. W. H. Stevens, Phys. Rev. B \textbf{27}, 2899 (1983).

[27] M. D. Kaplan and B. G. Vekhter, \textit{Cooperative Phenomena in
Jahn-Teller Systems} (Plenum, New York, 1995), p. 112.

[28] A. V. Babinskii \textit{et al.}, Pis'ma Zh. Eksp. Teor. Fiz \textbf{57},
289 (1993).

\bigskip

Figure captions:

Figure 1: (a) The PDF corresponding to the local atomic structure for x = 0.1
at representative temperatures is shown in the region of the Co-O peak. \ (b)
Split of the Co-O peak to short (SB) and long (LB) bonds at all measured temperatures.

Figure 2: The Co-O PDF peak is shown for 3 compositions at (a) RT and (b) 10
K. \ (c) The percent population is estimated for LS-Co$^{4+}$ assuming
localized charges to single sites, total LS sites that combine LS-Co$^{4+}$
and LS-Co$^{3+}$, and IS-Co$^{3+}$ ions in the PMI. \ Deviations from 100 \%
are a strong indication for the presence of the third state, that of
JT-Co$^{3+}$. (d) The composition dependence of the long (LB) and short (SB)
Co-O bonds plotted at RT and 10 K.

Figure 3: A plot of S$\left(  \omega\right)  $ determined for x = 0, 0.10,
0.30 and 0.40 at 10 K. \ Two gaussian functions (centered at $\sim$22 and 24
meV) are fit for 0, 0.1 and 0.3 and one for 0.4 (centered at $\sim$23
meV).\ The band formation around 22 meV is clearly separated by 0.3 and
signifies an unusual electron-lattice interaction in the metallic state. \ In
0.4 the two bands merge to ones.

Figure 4: The temperature dependence of S$\left(  \omega\right)  $ expanded
around 19-28 meV for 30 \% shows how the two modes become narrower through the IMT.

\end{document}